\documentclass{aa}
\usepackage{graphicx}
\usepackage{txfonts}
\usepackage{natbib}
\usepackage{longtable} 
\begin{document}
\title{X-ray flare in XRF 050406: evidence for prolonged engine activity} 
\author{P. Romano\inst{1}, 	
	A.~Moretti\inst{1}, 
	P.~L.~Banat\inst{1}, 
	D.~N.~Burrows\inst{2}, 
	S.~Campana\inst{1},  
	G.~Chincarini\inst{1},
	S.~Covino\inst{1},
	D.~Malesani\inst{3}, 
	G.~Tagliaferri\inst{1}, 
	S.~Kobayashi\inst{2,4}, 
	B.~Zhang\inst{5}, 
	A.~D.~Falcone\inst{2}, 
	L.~Angelini\inst{6,7}, 
	S.~Barthelmy\inst{6}, 	
	A.~P.~Beardmore\inst{8}, 
	M.~Capalbi\inst{9}, 
	G.~Cusumano\inst{10}, 
	P.~Giommi\inst{11}, 
	M.~R.~Goad\inst{8}, 
	O.~Godet\inst{8}, 
	D.~Grupe\inst{2},
	J.~E.~Hill\inst{6,12},
	J.~A.~Kennea\inst{2}, 
	V.~La~Parola\inst{10}, 
	V.~Mangano\inst{10},  
	P.~M{\' e}sz{\' a}ros\inst{2},
	D.~C.~Morris\inst{2}, 
	J.~A.~Nousek\inst{2}, 
	P.~T.~O'Brien\inst{8}, 
	J.~P.~Osborne\inst{8},
	A.~Parsons\inst{6}, 
	M.~Perri\inst{9},  
	C.~Pagani\inst{2,1}, 
	K.~L.~Page\inst{8}, 
 	A.~A.~Wells\inst{8}, 	
	N.~Gehrels\inst{6} 	
	}
\offprints{P.\ Romano}
\institute{INAF--Osservatorio Astronomico di Brera, Via E.\ Bianchi 46, I-23807 Merate (LC), Italy 
              \email{romano@merate.mi.astro.it}
  	\and Department of Astronomy \& Astrophysics, Pennsylvania State University,  
		525 Davey Lab, University Park, PA 16802, USA 
	\and International School for Advanced Studies (SISSA-ISAS), Via Beirut 2-4, I-34014 Trieste, Italy 
  	\and Center for Gravitational Wave Physics, Pennsylvania State University,
		104 Davey Lab, University Park, PA 16802, USA 
  	\and Department of Physics, University of Nevada, Las Vegas, NV 89154-4002, USA
	\and NASA/Goddard Space Flight Center, Greenbelt, MD 20771, USA  
	\and Department of Physics and Astronomy, The Johns Hopkins University,
	    3400 North Charles Street, Baltimore MD 21218 USA
	\and X-Ray and Observational Astronomy Group, Department of Physics \& Astronomy, University of Leicester, LE1 7RH, UK
  	\and ASI Science Data Center, via G.\ Galilei, I-00044 Frascati (Roma), Italy 
  	\and INAF--Istituto di Astrofisica Spaziale e Fisica Cosmica Sezione di Palermo,
              Via U.\ La Malfa 153, I-90146 Palermo, Italy  
	\and Agenzia Spaziale Italiana, Unit\`a Osservazione dell'Universo, Viale Liegi 26, I-00198 Roma, Italy 
	\and Universities Space Research Association, 10211 Wincopin Circle, Suite 500, Columbia, MD, 21044-3432, USA
	}  
\date{Received: 2005 September 7; accepted 2006 January 3}
\abstract{
We present observations of XRF 050406, the first burst detected by Swift
showing a flare in its X-ray light curve. 
During this flare, which peaks at $t_{\rm peak} \sim 210$\,s after the BAT trigger, 
a flux variation of $\delta F / F \sim 6$ in 
a very short time $\delta t / t_{\rm peak} \ll 1$ was observed. 
Its measured fluence in the 0.2--10 keV band was $\sim 1.4 \times 10^{-8}$ erg cm$^{-2}$, 
which corresponds to 1--15\% of the prompt fluence.
We present indications of spectral variations during the flare. 
We argue that the producing mechanism is late internal shocks,
which implies that the central engine is still active at $210$\,s,
though with a reduced power with respect to the prompt emission. 
The X-ray light curve flattens to a very shallow slope with decay index 
of $\sim 0.5$ after $\sim 4400$\,s, which also supports continued central 
engine activity at late times.
This burst is classified as an 
X-ray flash, with a relatively low fluence ($\sim 10^{-7}$ erg cm$^{-2}$ in 
the 15--350 keV band, 
$E_{\rm iso} \sim 10^{51}$ erg), 
a soft spectrum (photon index 2.65), no significant
flux above $\sim 50$ keV and a peak energy $E_{\rm p} < 15$ keV. 
XRF 050406 is one of the first examples of a well-studied X-ray light curve of an XRF. 
We show that the main afterglow characteristics are qualitatively similar 
to those of normal GRBs. 
In particular, X-ray flares superimposed on a power-law light curve have now been 
seen in both XRFs and GRBs. This indicates that a similar mechanism may be at work 
for both kinds of events. 

}
\authorrunning {P.\ Romano}
\titlerunning {X-ray flare in XRF 050406}
\maketitle
	\section{Introduction\label{grb050406:introd}}
The Swift Gamma-Ray Burst Explorer (\citealt{SWIFT})
was successfully launched on 2004 Nov 20.
Its payload includes one wide-field instrument, the gamma-ray (15--350 keV) 
Burst Alert Telescope (BAT; \citealt{BAT}),  
and two narrow-field instruments, 
the X-Ray Telescope (XRT; \citealt{XRT2}) 
and the Ultraviolet/Optical Telescope (UVOT; \citealt{UVOT}).
BAT detects the bursts, calculates their position 
to $\sim 1$--$4 \arcmin$ accuracy and triggers an autonomous 
slew of the observatory to point the two narrow-field instruments. 
The XRT, which operates in the 0.2--10\,keV energy range, 
can provide $\sim 5\arcsec$ positions, 
while the UVOT, which operates in the 1700--6000\,\AA{} wavelength range, 
can further refine the afterglow localization to $\sim 0\farcs5$. 
With its unique fast re-pointing capabilities Swift set out to investigate 
the very early phases of gamma-ray burst (GRB) afterglows, 
beginning as early as one minute after the BAT trigger. 
During the initial activation and calibration phases, which ended on 2005 Apr 5, 
Swift discovered 25 GRBs. 
The narrow-field instruments were re-pointed towards seven of them within a 
few hundred seconds, and such is the case for GRB 050406.

On 2005 Apr 6 at 15:58:48.40 UT, BAT triggered on GRB 050406 
(trigger 113872; \citealt{GCN3180}), and located it at 
RA(J2000$) = 02^{\rm h}  17^{\rm m}  53^{\rm s}$, 
Dec(J2000$) =-50^{\circ}  10\arcmin  52\arcsec$, 
with an uncertainty of 3 arcmin (95\% containment; \citealt{GCN3183}).
The derived value for the time during which 90\% of the burst fluence is observed 
was $T_{90}=5\pm1$\,s in the 15--350 keV band. 
In the 15--25 keV band the light curve peak had a fast-rise, 
exponential decay (FRED) profile, while in the 25--50 keV band, the shape was more 
symmetric, with the peak starting $\sim 2$\,s earlier (\citealt{GCN3183}). 
Both the peak and time-averaged spectra were well fit by a simple power-law
with a time-averaged spectrum photon index of 2.38$\pm$0.34 
(90\% confidence; \citealt{GCN3183}). 
The fluence in the 15--350 keV band was $9.0 \times 10^{-8}$ erg cm$^{-2}$.
The gamma-ray characteristics of this burst, i.e.\ 
the softness of the observed spectrum and 
the absence of significant emission above $\sim 50$ keV, 
classify GRB 050406 as an X-ray flash (XRF; \citealt{Heiseea01}). 
From now on, we shall therefore refer to this event as XRF 050406. 

Swift executed a prompt slew. 
The XRT imaged the BAT field only 84\,s after the trigger 
but no bright X-ray source could be detected within the field of view. 
However, a refined on-ground analysis revealed a previously uncatalogued  
X-ray source (\citealt{GCN3181,GCN3184}).
From the very first examination of the down-linked data it was clear that the 
afterglow of this burst was peculiar. Indeed, after an initial decay, the 
X-ray count rate began rising, peaking at $\approx 220$\,s, and subsequently 
decaying again (\citealt{GCN3184}).

Ground-based observations started as soon as the burst discovery was reported via the 
GCN network. 
The Magellan/Clay telescope imaged the XRT error circle
with LDSS-3 in the $R$ and $i$ bands and found a single faint source ($R=22.0\pm0.09$ mag, 
7.8 hr after the burst) located at 
RA(J2000$)=02^{\rm h} 17^{\rm m} 52\fs3$, Dec(J2000$)=-50^{\circ} 11\arcmin 15\arcsec$ 
with an uncertainty of $\sim 0\farcs5$ in each coordinate (\citealt{GCN3185,Bergerea05b}).
Similarly to XRT, UVOT also imaged the field at the end of the slew 
(starting from $\sim 88$ s after the trigger) and though it failed to detect the afterglow on-board
(\citealt{GCN3182}), 
subsequent on-ground analysis revealed a source within the XRT error circle at the 
4.3- (19.0 mag), 3.0- and 2.5-$\sigma$ detection levels in the $U$, $B$ and $V$ bands, 
respectively. 
The UVOT position was RA(J2000$)=02^{\rm h} 17^{\rm m} 52\fs2$, 
Dec(J2000$)=-50^{\circ} 11\arcmin 15\farcs8$, consistent with the 
Magellan one.
By the time the second UVOT observation (1.3 hr later) was performed, 
the source was not detected in the $U$ band, confirming it  as the 
afterglow of XRF 050406. 
\citet{Schady06} obtained 
an estimate of $z=2.44\pm0.36$ from fitting the broad band spectrum (combined UVOT and XRT data).

In this paper we present observations of the first Swift burst  
where a flare is clearly detected in its X-ray light curve, 
during which the source count rate increased by a factor of $\ga 6$.
This feature had never been observed before in Swift data, and 
had rarely been observed before in any X-ray afterglow (\citealt{piroea05}). 
This paper is organized as follows. 
In Sect.\ \ref{grb050406:dataredu} we describe our observations and data reduction; 
in Sect.\ \ref{grb050406:dataanal} we describe our spatial, timing and spectral data analysis; 
in Sect.\ \ref{grb050406:discussion} we discuss our findings. 
Finally, in Sect.\ \ref{grb050406:conclusions} we summarize our conclusions. 
Throughout this paper the quoted uncertainties are given at 90\% confidence level 
for one interesting parameter (i.e., $\Delta \chi^2 =2.71$) unless otherwise stated.
Times are referred to the BAT trigger $T_0$, $t=T-T_0$.
The decay and spectral indices are parameterized as follows, 
$F(\nu,t) \propto t^{-\alpha} \nu^{-\beta}$, where $F_{\nu}$ (erg cm$^{-2}$ s$^{-1}$ Hz$^{-1}$) is the 
monochromatic flux as a function 
of time $t$ and frequency $\nu$; we also use $\Gamma = \beta +1$ as the photon index, 
$N(E) \propto E^{-\Gamma}$ (ph keV$^{-1}$ cm$^{-2}$ s$^{-1}$).

	\section{Observations and data reduction\label{grb050406:dataredu}} 
	
\subsection{BAT observations\label{grb050406:batobs}} 

Table~\ref{grb050406:tab_obs} reports the log of the observations that were used for this work.  
The BAT data were analyzed using the standard BAT analysis software distributed 
within FTOOLS v6.0. 
The burst is detected in the first two standard bands (15--25 and  25--50\,keV)
while virtually no flux is observed above 50 keV.
We find $T_{\rm 90} = 6.1\pm 1.0$\,s in the 15--150 keV band.

The BAT spectra were extracted over the full time interval over which the burst was 
detected ($T_{\rm tot}$), in the interval covering the 1-s peak $T_{\rm peak}$,  
and for the  $T_{\rm 90}$ and $T_{\rm 50}$ intervals. 
Response matrices were generated with the task {\tt batdrmgen} using the latest 
spectral redistribution matrices. 
For our spectral fitting (XSPEC v11.3.2) we considered the 15--150 keV energy range. 
All spectra are well fit with a simple power law with photon index 
$\Gamma_\gamma \sim 2.65$ 
(see details in Table~\ref{grb050406:tab_specfits}). 
There is no evidence of a spectral break within the BAT energy range, thus 
constraining the peak energy $E_{\rm p}<15$\,keV. 
The indices are steeper (softer) although consistent with the ones reported by 
\citet{GCN3183}, due to the different energy ranges used for the spectral fitting. 
No significant improvements are found using either a cutoff power-law or a Band model 
(\citealt{Band}).
The 1-s peak photon flux was 
 $(2.3_{-0.4}^{+2.8})\times 10^{-8}$  erg cm$^{-2}$ s$^{-1}$ (15--350 keV band), while 
the fluence was $\mathcal{F} = (1.0 ^{+1.13}_{-0.36})\times 10^{-7}$ erg cm$^{-2}$ (15--350 keV band). 
This fluence corresponds to an isotropic-equivalent energy 
$E_{\rm iso} = (1.4^{+1.6}_{-0.6})\times 10^{51}$ erg
(in the rest frame 52--1204 keV) assuming $z=2.44\pm0.36$ (\citealt{Schady06}).

\subsection{XRT observations\label{grb050406:xrtobs}}

	\begin{figure}
 	 	\resizebox{\hsize}{!}{\includegraphics[angle=270]{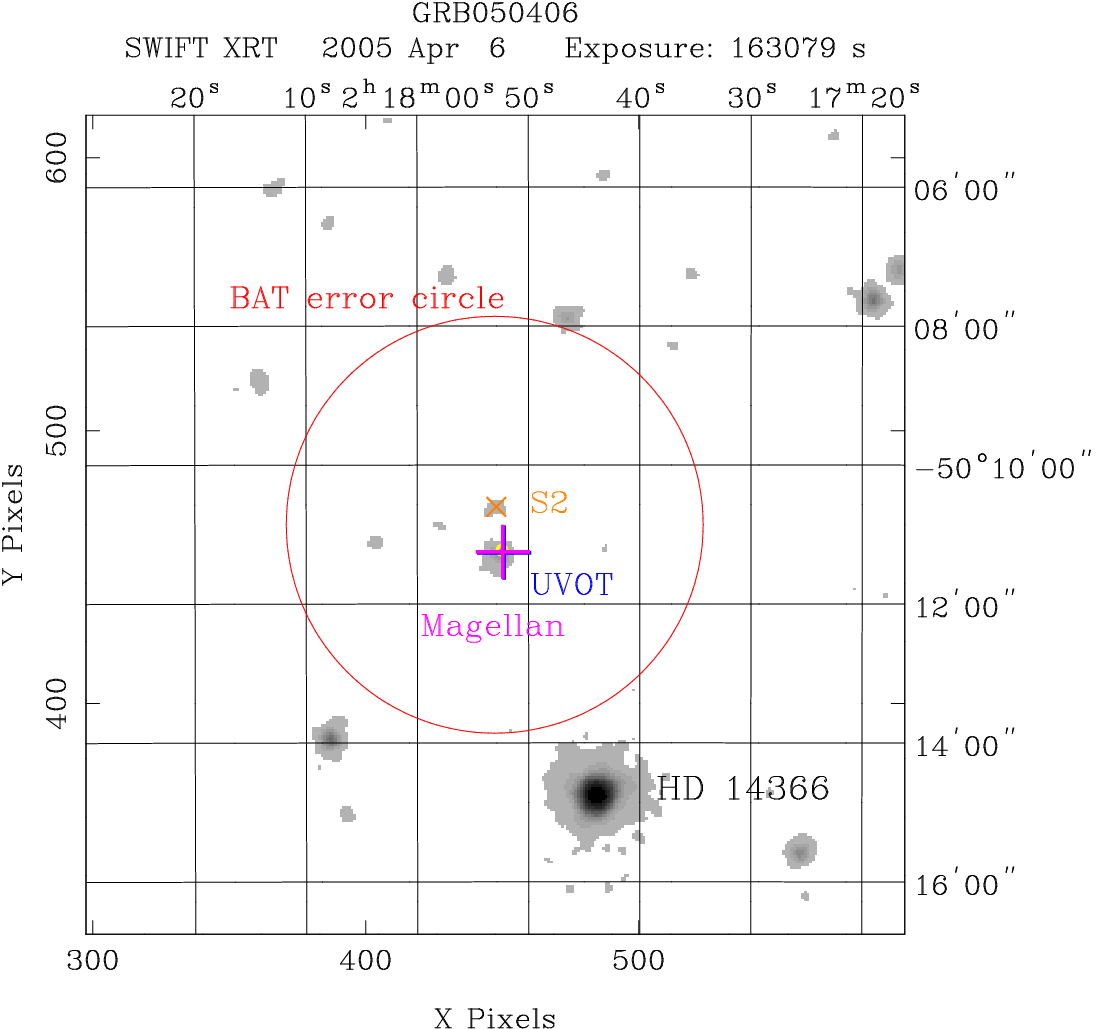}}  
 		\caption{XRT image of XRF 050406, 
		obtained from the total $\sim$163\,ks PC mode data.
		The field is centred on the 3$\arcmin$ radius BAT error circle. 
		Also shown is the XRT 4\farcs2 error circle, as well as
	       	the Magellan (\citealt{GCN3185}) and UVOT (\citealt{GCN3186}) 
		optical counterpart positions; 
		the optical points are so close they cannot be distinguished on this scale.
		S2 is a serendipitous source located at 
		RA(J2000$)=02^{\rm h} 17^{\rm m} 52.^{\rm s}9$, Dec(J2000$)=-50^{\circ} 10\arcmin 36\farcs1$.
		}
 		\label{grb050406:fig_map}
	\end{figure}

In order to cover the dynamic range and rapid variability expected from GRB afterglows 
and to provide rapid-response, automated observations, XRT was designed to support different 
readout modes that optimize the collected information as the flux of the burst diminishes. 
The XRT supports four major readout modes, one imaging (IM), 
two timing, Piled-up/Low-rate Photodiode (PuPD and LrPD) and Windowed Timing (WT), 
and one Photon-Counting (PC). A detailed description of XRT modes can be found in \citet{xrtmodes}.
In the nominal operating state the mode switching is based on the source flux and 
is fully automated (auto state) to minimize pile-up in the data. 

The XRT observations of XRF 050406 started on 2005 Apr 6 at 16:00:12 UT, 
only 84\,s after the trigger, and ended on 2005 Apr 22, thus summing up a 
total net exposure (in PC mode) of $\sim 163$ ks spread over a $\sim$16\,d baseline.
The monitoring is organized in 9 observations (000, 001, 002, 005, 006, 008, 009, 010, 011) 
and 183 snapshots (continuous pointings at the target). 
This was the first burst to occur after the formal end of the calibration phase (2005 Apr 5), and 
the first (000) observation was performed as an automated target (AT) with XRT in auto state. 
Therefore, during observation 000 the automated mode switching made XRT take an initial 2.5\,s 
image (IM at $t=84$\,s), immediately followed by one PuPD ($t=90$\,s) and one LrPD ($t=91$\,s) frame. 
Then at $t=92$\,s a series of 5 WT frames was taken until the on-board measured count rate
was low enough for XRT to switch to PC mode ($t=99$\,s). 
After this, XRT repeatedly switched between WT and PC modes because of an increased 
background level (see below). 
Since the signal-to-noise (S/N) in these late WT frames is low, we did not include 
them in our analysis (Table~\ref{grb050406:tab_obs}).

The XRT data were first processed by the Swift Data Center at NASA/GSFC into
Level 1 products (calibrated and quality-flagged event lists). Then 
they were further processed with the XRTDAS (v1.4.0) software package written by the
Agenzia Spaziale Italiana (ASI) Science Data Center and distributed within
FTOOLS v6.0 to produce the final cleaned event lists. We ran the
task {\tt xrtpipeline} (v0.8.8) applying standard filtering and screening criteria,
i.e., we cut out temporal intervals during which the CCD temperature was higher
than $-47$ $^\circ$C, and we removed hot and flickering pixels. These are present
because the CCD is operating at a temperature higher than the design
temperature of $-100$ $^\circ$C due to a failure in the active cooling system. 
An on-board event threshold of $\sim$0.2 keV (un-reconstructed pulse-height PHAS[1]$>80$) 
was also applied to the central pixel, which has
been proven to reduce most of the background due to either the bright Earth
limb or the CCD dark current (which depends on the CCD temperature). 
These two sources of background are the main reason for the switching between PC and WT mode
even when the source count rate is below 1 counts s$^{-1}$. 
	
Throughout the monitoring campaign the CCD temperature was $<-50$ $^\circ$C, 
with the exception of part of observations 002 and 005, 
where it became as high as $-43.5$ and $-45$ $^\circ$C, respectively; 
those data were therefore screened out.  
For our analysis we further selected XRT grades 0--12 and 0--2 for PC and WT data, respectively
(according to Swift nomenclature; \citealt{XRT2}).

	\section{Data analysis\label{grb050406:dataanal}} 

\subsection{Spatial analysis\label{grb050406:spatial}} 

Figure~\ref{grb050406:fig_map} shows the 163 ks XRT image accumulated in PC mode in the 
0.2--10 keV energy band. 
We detected two previously uncatalogued sources within 1 arcmin of the optical 
burst coordinates. 
The brightest uncatalogued source, which we identified as the fading X-ray 
counterpart of the burst, is present in the first four XRT snapshots.
The source is piled-up during the initial 500\,s of PC data. 
Therefore, to obtain an unbiased position, we rely on the 
remainder of the PC data in the first observation, which has a 
net exposure of 49.8\,ks.  
We used the {\tt xrtcentroid} task (v0.2.7) and found that the 
afterglow position is RA(J2000$)=2^{\rm h} 17^{\rm m} 52\fs4$, 
Dec(J2000$)=-50^{\circ} 11\arcmin 13\farcs6$. 
We estimate its uncertainty to be 4\farcs2 (90\% confidence level). 
This position takes into account the correction for the misalignment
between the telescope and the satellite optical axis (\citealt{centroids}). 
Figure~\ref{grb050406:fig_map} shows the XRT error circle, as well as the 
3\arcmin{} BAT error circle (\citealt{GCN3183}; 95\% containment) and the 
optical counterpart coordinates determined by Magellan (\citealt{GCN3185}) 
and by UVOT (\citealt{GCN3186}). 
The XRT coordinates are 23\arcsec{} from 
the BAT ones, and 1\farcs6 and 2\farcs8 from the Magellan and UVOT ones, respectively. 
XRF 050406 was detected ({\tt XIMAGE} v4.3) in the first four snapshots individually, 
but not from the 5th on. 
The second source, S2, is located at RA(J2000$)=02^{\rm h} 17^{\rm m}  52\fs9$,  
Dec(J2000$)=-50^{\circ} 10\arcmin 36\farcs1$ and has a constant rate
($3.8\pm 0.7) \times 10^{-4}$ counts s$^{-1}$ throughout the observation 
campaign. 

	\begin{figure*}
	 	\resizebox{\hsize}{!}{\includegraphics{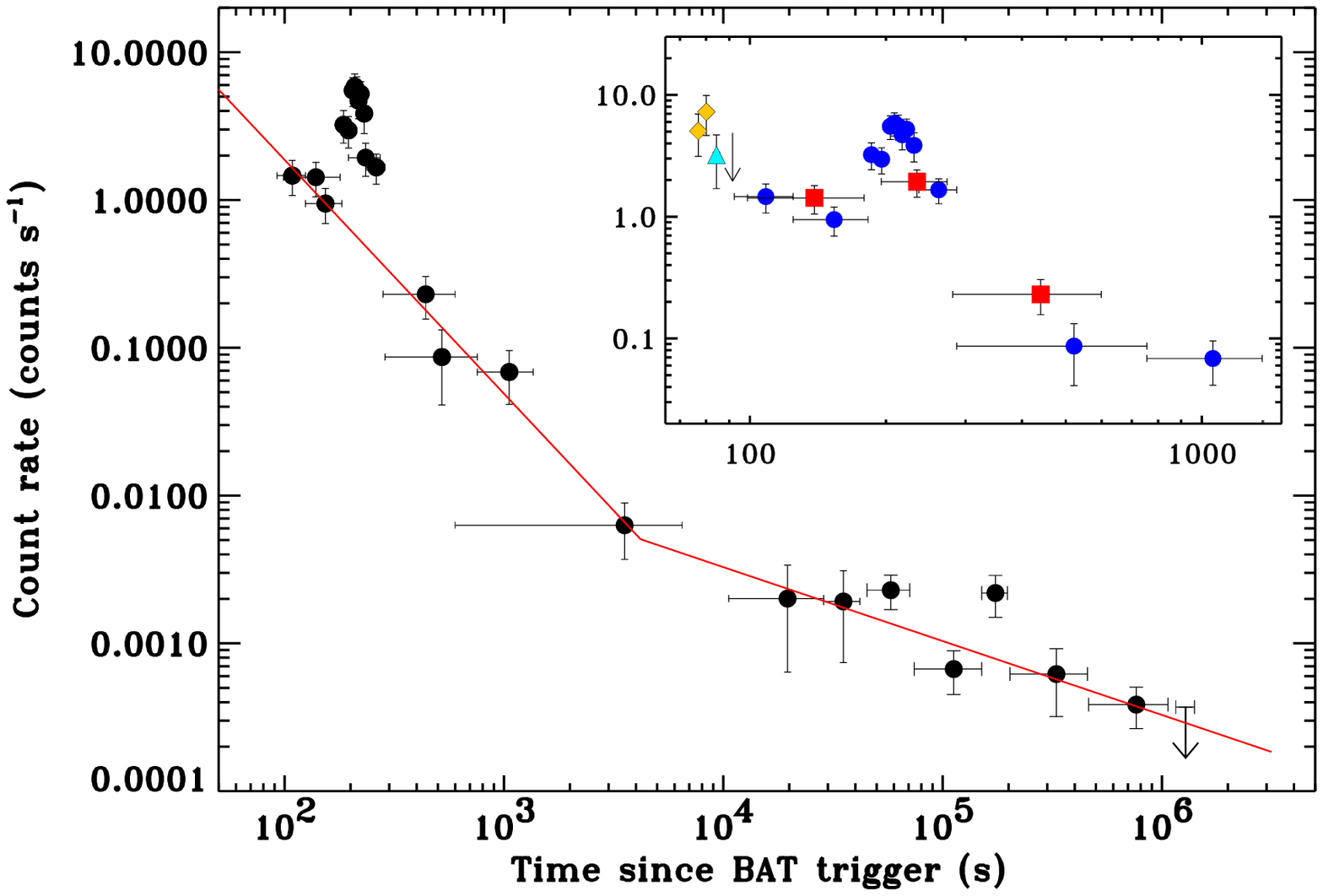}}
 		\caption{X-ray light curve of the XRF 050406 afterglow in the 0.2--10\,keV energy band. 
		The curve is background-subtracted and the time is referred to the BAT trigger, 2005 Apr 06 at
		15:58:48.4 UT (\citealt{GCN3180}). The last point after 10$^6$\,s is a 3-$\sigma$ upper limit. 
		{\bf Inset}: details of the first $\sim$ 1000\,s, which include data in all XRT modes. 
		The (yellow) diamonds represent LrPD mode data taken during the latter portion of the slewing phase; 
		the (cyan) triangle is the initial IM point (84\,s after the trigger, see Table~\ref{grb050406:tab_obs}), 
		the downward-pointing arrow is a LrPD limit (pointing, 91\,s after the trigger), 
		the (blue) circles are WT mode data (starting from 92\,s after the trigger), 
		and the (red) squares are PC mode data (starting from 99\,s after the trigger).
		The data have been corrected for pile-up (where appropriate) and PSF losses. 
		The solid (red) line represents the best-fit broken power-law model to the light curve 
		(excluding the flare).
		}
 		\label{grb050406:fig_lcvs}
	\end{figure*}

\subsection{Temporal analysis\label{grb050406:timing}} 

During the first 500\,s of the XRT observation the intensity of the afterglow was high enough 
to cause pile-up in the PC mode data. To account for this effect we extracted 
the source events within an annulus with a 30-pixel outer radius ($\sim71$\arcsec) 
and a 2-pixel inner radius. 
These values were derived by comparing the observed and nominal PSF. 
For the PC data collected after the first 500\,s, the entire circular region (30-pixel radius) 
was used, instead. 
In both cases we further disregarded data within a circular region centred on the serendipitous 
source S2 (which lies within the 30-pixel PC source extraction region) 
with a 7.17 pixel (17\arcsec) radius. 
The WT data were extracted in a rectangular region 31 pixels long along the image strip 
(and 20 pixels wide), which excludes the data from the source S2. 
The selected extraction regions correspond to $\sim 69$\,\% (piled-up PC), 
$\sim 95$\,\% (non piled-up PC), 
and $\sim 94$\,\% (WT) of the XRT PSF. 
To account for the background, data were also extracted in PC mode within 
a circular region (radius $130\arcsec =54.8$ pixels) 
and in WT mode within a rectangular box (40$\times$20 pixels), 
in locations far from background sources. 
The mean PC background in the 0.2--10 keV band was found to be constant throughout the observations 
and, normalized to the PC source extraction region, it had a value of $\sim 2.6 \times 10^{-3}$ counts s$^{-1}$.
The mean WT background in the same energy band and normalized to the WT source region was 
$\sim 4.6 \times 10^{-2}$ counts s$^{-1}$. 

Figure~\ref{grb050406:fig_lcvs} shows the background-subtracted light curve extracted in the 
0.2--10 keV energy band, with the BAT trigger as origin of time. 
We considered WT data for the first snapshot of the first observation, 
and PC data for all 9 available observations (see Table~\ref{grb050406:tab_obs}). 
During the initial phases of the afterglow evolution ($t< 4\times10^4$\,s) 
we binned the source counts with a minimum of 30 counts per time bin, and dynamically subtracted the 
normalized background counts in each bin. The PC mode data were corrected for the effects of pile-up. 
We note that, by keeping to the minimum number of counts per time bin criterion, 
we created several bins during the first snapshot, but subsequently needed to merge data belonging 
to snapshots 1 and 2 (point at $\sim 4$\,ks), then from snapshots 3 and 4 (point at $\sim 20$\,ks), 
and later on from snapshots 5 through 8 (point at $\sim 35$\,ks). 
Afterwards, we used {\tt XIMAGE} with the option  {\tt SOSTA}, which 
calculates vignetting- and PSF-corrected count rates within a specified box, 
and the background in a user-specified region. 
To ensure uniformity with the early light curve, the background was estimated in the same 
region as the one used for the initial part of the light curve. 
We thus obtained a signal-to-noise ratio S/N $\ga 3$ (the only exception being the point at 
$\sim 33$ ks which has S/N $\ga 2$). The last point is a 3-$\sigma$ upper limit. 
This latter method is preferred for the construction of the late part of the light curve
since it better accounts for the background in a low-counts regime. 
We note, however, that extracting the light curve in the same 30-pixel source region up to the 
end of the last observation, we obtained  fully consistent results, albeit with a noisier 
light curve. 
We also note that the residual contribution of the serendipitous source S2 within the source 
extraction region is $\la 19\%$ of the S2 counts, which corresponds to 
$\la (7\pm1)\times 10^{-5}$ counts s$^{-1}$. 
Therefore, S2 only makes a marginal contribution to the afterglow light curve, which amounts to 
$<$20\% of the last point.

The light curve clearly shows a complex behaviour, with a power law decay underlying 
a remarkable flare which peaks at $\approx 210$ s after the BAT trigger 
(see Fig.~\ref{grb050406:fig_lcvs}, inset). 
To fit the light curve we used the BAT trigger as reference time
and we only considered spectroscopic-mode data obtained while XRT was pointing, 
thus excluding the early LrPD, the LrPD upper limit and the IM point. 
Further excluding the data taken during the flare (180\,s $<t<300$\,s), 
a fit with a simple power law yields $\chi^2_{\rm red}=4.32$ (12 degrees of freedom, d.o.f.), 
which is unacceptable. 
A fit with a broken power law $F(t) = K t^{-\alpha_1}$  for $t<t_{\rm b}$ and  
                    $F(t) = K\,t_{\rm b}^{-\alpha_1} \, (t/t_{\rm b})^{-\alpha_2}$ for $t>t_{\rm b}$,  
where $t_{\rm b}$ is the time of the break, yields 
$\alpha_1=1.58^{+0.18}_{-0.16}$ and $\alpha_2=0.50\pm0.14$, and a break at $\sim 4200$ s after the
BAT trigger.
This latter model yields a good fit ($\chi^2_{\rm red}=1.20$, 10 d.o.f.), a significant improvement 
over the simple power law (null hypothesis probability $=1.7\times 10^{-3}$, equivalent to 3.2 $\sigma$),
but some of the parameters are not well constrained.
Alternatively, a fit with two smoothly joined power laws
$F(t) = K^\prime [(t/t_{\rm b})^{-\alpha_1} + (t/t_{\rm b})^{-\alpha_2}]$
yields $\chi^2_{\rm red}=1.29$  (10 d.o.f.) with similar values for the inferred parameters.
A summary of the fits to the light curve can be found in Table~\ref{grb050406:tab_lcvfits}.
As a reference, the 0.2--10 keV unabsorbed 
flux at $t_{\rm b}$ is $(4\pm1)\times 10^{-13}$  erg cm$^{-2}$ s$^{-1}$
(we adopted a count rate to unabsorbed flux conversion factor of 
$6.5 \times 10^{-11}$ erg cm$^{-2}$ count$^{-1}$,
obtained from the best fit models derived in Sect.~\ref{grb050406:spectral})
and the luminosity in the 0.7--34.4 keV band is $(1.9\pm0.9)\times 10^{46}$ erg s$^{-1}$.  

During the flare a rebrightening of the source by a factor of $\ga 6$ in flux was observed 
between $t \sim 154$\,s and the peak at $\sim 210$\,s.
Both the rising and the falling part of the flare had very steep slopes that, 
when fit with a simple power law, yield $\alpha_{\rm 1,flare}=-5.8^{+1.6}_{-2.1}$ 
and $\alpha_{\rm 2,flare}=6.7\pm1.0$. When the underlying power-law afterglow is subtracted, 
the fit yields $\alpha_{\rm 1,flare}=-6.8^{+2.4}_{-2.1}$ and $\alpha_{\rm 2,flare}=6.8^{+3.6}_{-2.0}$ 
and the peak is at $213\pm7$\,s from the BAT trigger. 
In all cases the errors are dominated by the uncertainty in the placement of the flare boundaries.
The flare can also be characterised, as a simple parametric description, 
as a Gaussian line. A combined broken power law and Gaussian model fit yields a 
peak at $211.1^{+5.4}_{-4.4}$\,s 
($61.4^{+1.6}_{-1.3}$\,s in the rest frame) 
and a width $17.9^{+12.3}_{-4.6}$\,s 
($\chi^2_{\rm red}=1.58$, 17 d.o.f.). 
In this case the ratio of the characteristic time-scale and the peak time is 
 $\delta t / t_{\rm peak} \sim 0.08$ or 0.20, when using the Gaussian width or its 
FWHM ($42.2^{+29.0}_{-10.8}$\,s), respectively. In either case,  
$\delta t / t_{\rm peak} \ll 1$, which puts severe constraints on the
emission mechanisms that can produce the flare. 
We shall address this issue in the discussion section. 
Integration of the Gaussian best-fitting function yields 
an estimate of the fluence of the flare, $(1.4 \pm 1.0) \times 10^{-8}$ erg cm$^{-2}$, 
corresponding to an energy of $(2.0 \pm 1.4) \times 10^{50}$ erg.  
The large error reflects the uncertainty on the actual model used 
for the integration of the flare. 

\begin{figure}
	 	\resizebox{\hsize}{!}{\includegraphics{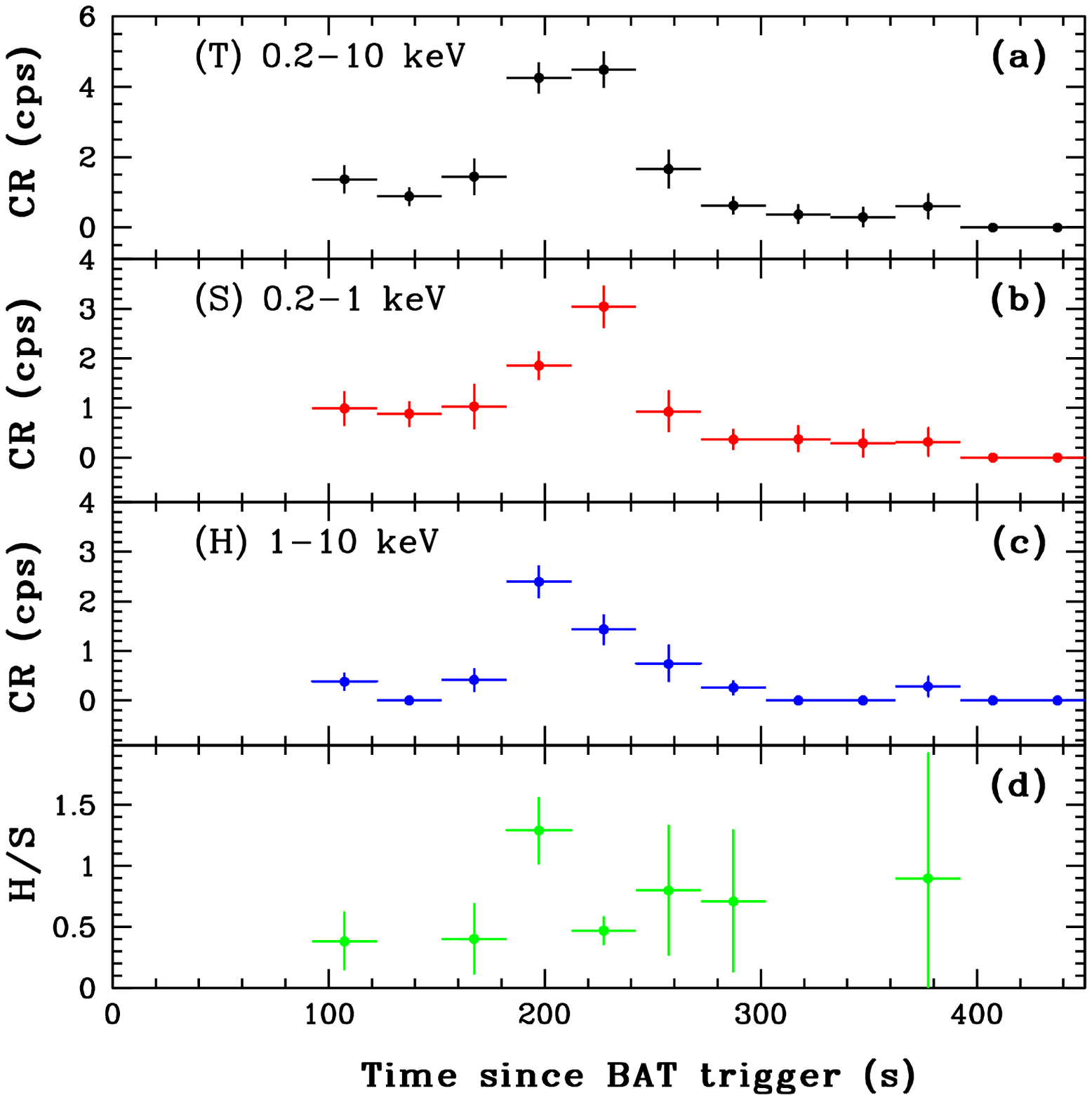}}
 		\caption{WT background-subtracted light curves. 
		{\bf (a)}: Total band  ({\bf T}, 0.2--10 keV). 
		{\bf (b)}: Soft band  ({\bf S}, 0.2--1 keV).
		{\bf (c)}: Hard band  ({\bf H}, 1--10 keV). 
		{\bf (d)}: Ratio of hard to soft count rates. 
		}
 		\label{grb050406:fig_hr}
	\end{figure}

We also extracted events from the first snapshot WT data in two more energy bands, 
0.2--1 keV (soft, S) and 1--10 keV (hard, H), as well as the total band,
0.2--10 keV. We used the same regions as the ones described above, 
a constant time binning of 30\,s and dynamically subtracted their respective backgrounds.  
Figure~\ref{grb050406:fig_hr} shows the three background-subtracted light curves, 
as well as the ratio H$/$S.
Indeed, during the rising portion of the flare the hard band flux increases by a factor of 
$\ga 6$ while the soft band flux only increases slightly, 
so that the spectrum of the flare starts off harder than the underlying afterglow, 
and then evolves into a softer state as its flux decreases; 
this can be seen in the following time bin, when the soft band flux peaks with a flare to 
pre-flare flux ratio of $\sim 3.5$. 
This yields an indication of spectral evolution during the flare
as a $\sim 3$-$\sigma$ excess over a constant fit to H$/$S.
It should be noted that this behaviour is reminiscent of that observed in the 
{\it prompt} emission (\citealt{Fordea95}),
 with the harder band peak preceding the softer band peak. 

At $t \sim 1.7 \times 10^{5}$\,s a second faint bump is observed.
Its significance is not high, since it is detected as a 2-$\sigma$
excess over the underlying afterglow. Similar late-time bumps 
have been observed in other Swift-detected GRBs 
(e.g.\ GRB 050502B; \citealt{Falconeea05}).

	\subsection{Spectral analysis\label{grb050406:spectral}} 

The afterglow of XRF 050406 was very faint, hence it is not possible to perform 
time-resolved spectroscopy to distinguish the spectral properties of the 
afterglow proper from the ones of the flare observed in the light curve. 
Therefore, we proceeded as follows. 
Spectra of the source and background were extracted in the regions described in 
\S\ref{grb050406:timing} from the first observation (000) event files.  
PC and WT spectra were extracted during the first $\sim$ 500\,s 
of the PC observation (see Table~\ref{grb050406:tab_specfits} for times referred to 
$T_0$), when PC data are piled-up and when the flare is observed in the light curve. 
We also extracted PC spectra after the first 500\,s during the
first 4 snapshots. For the latter we used a circular region with a 10 pixel radius 
(corresponding to $\sim 80$\,\% of the XRT PSF) to minimize the background 
and to be able to use the Cash statistics (\citealt{Cash79}). 
Ancillary response files (ARF) were generated with the task {\tt xrtmkarf} 
within FTOOLS v6.0 using the latest ARF distribution (v003). 
These ARFs account for different extraction regions and PSF corrections. 
We used the latest spectral redistribution matrices (v007). 
The energy ranges adopted for spectral fitting were 0.5--10 keV 
and 0.2--10 keV for WT and PC, respectively. 

We first performed a fit with an absorbed ({\tt wabs} in XSPEC) power law 
to the WT data (166 counts), 
which were rebinned with a minimum of 20 counts per energy bin to 
allow $\chi^2$ fitting within XSPEC. 
The Hydrogen column was initially kept as a free parameter, 
and then frozen to the Galactic value ($N_{\rm H}^{\rm G}= 2.8 \times 10^{20}$ cm$^{-2}$, \citealt{DL90}) 
when the fit yielded a value lower than (although consistent with) the Galactic one. 
The fit was good, $\chi^2_{\rm red}=1.0$ for 6 d.o.f., 
and yielded $\Gamma=2.11_{-0.28}^{+0.31}$.
We then performed a fit with the same model to the remainder 
of the PC data during snapshots 1 through 4 
using Cash statistics which is more appropriate given the low number of counts 
(21 un-binned counts) and calculated the goodness of the fit via $10^4$ Montecarlo 
simulations. The fit was good and yielded consistent results. 
We also performed simultaneous fits to the WT and PC (60 counts) spectra extracted during the first 
$\sim 500$\,s (using $\chi^2$ statistics) and of the PC data alone (using Cash statistics),
also obtaining consistent results. 
Table~\ref{grb050406:tab_specfits} summarizes the results of the fits.
We note that, given the current goodness of the XRT calibration 
(5\% systematic uncertainty for all observing modes and grade selections in the 
0.5--10 keV range; e.g., \citealt{Romanospie05}), an excess of  
$N_{\rm H}$ cannot be excluded and we find a 3-$\sigma$ upper limit to the
total (Galactic plus intrinsic) Hydrogen column along the line of sight of 
$N_{\rm H} < 9 \times 10^{20}$ cm$^{-2}$. 

We can therefore conclude that, during the first 600\,s after the burst, 
which include the X-ray flare observed in the light curve, the mean photon index is 
$\Gamma=2.1\pm0.3$, and that the photon index does not vary after the end of the flare.
However, we do have clues regarding the presence of spectral evolution during the flare
coming from the hardness ratio analysis (Sect.~\ref{grb050406:timing}), 
even though the statistics are not high enough to show it in the spectral analysis. 
As we will discuss later (Sect.~\ref{grb050406:disc_flares}), other afterglows with
larger amplitude X-ray flares demonstrate a strong spectral evolution 
of the flares.

	\section{Discussion\label{grb050406:discussion}} 

\subsection{Gamma-ray properties: similarity of XRFs and GRBs\label{grb050406:disc_prompt}}

The duration of this burst ($T_{\rm 90} = 5\pm 1$\,s 
in the 15--350 keV band) places this burst in the short  
tail of the long GRB population (\citealt{Kouveliotouea93}).  
Its fluence is relatively low ($1.0 \times 10^{-7}$ erg cm$^{-2}$ in the 
15--350 keV band) but not unusually faint. 
The gamma-ray characteristics of this burst are consistent with a classification  
as an X-ray flash (\citealt{Heiseea01}), or as an ``X-ray rich GRB'' (XRR). 
The softness of the observed spectrum, which is well fit in the 15--150 
keV band with a simple power law with photon index $\Gamma_\gamma=2.65$, 
and with no significant emission observed above $\sim 50$ keV, 
implies that the peak energy is below the BAT bandpass ($E_{\rm p} < 15$ keV). 
The operational definition of XRFs/XRRs (e.g.\ \citealt{Lambea04}) is
of a fast transient X-ray source characterized by a softness ratio 
SR$=\log [\mathcal{F} ({\mbox{2--30\,keV}})/\mathcal{F} ({\mbox{30--400\,keV}})] > 0$ for an XRF and 
$-0.5 < $SR$ < 0$ for an XRR. 
Extrapolation of the BAT spectrum, with the assumption of  
$E_{\rm p} < 2$ keV,  yields SR$=0.8^{+0.5}_{-0.4}$, which classifies this burst as an XRF. 
However, a break in the spectrum may well be present in the 2--15 keV band. 
In the most conservative case, i.e. assuming no flux below 15 keV, this event 
would be an XRR GRB, with SR$ = -0.2^{+0.2}_{-0.3}$.

The isotropic-equivalent gamma-ray energy of this event is 
$E_{\rm iso} = (1.4^{+1.6}_{-0.6})\times 10^{51}$ erg (Sect.~\ref{grb050406:batobs}), 
and this effectively puts XRF~050406 in the low-energy tail of GRB energies 
(\citealt{Bloomea03b}). Assuming that the Amati relation (\citealt{Amatiea02}) holds, 
we can infer a rest-frame $E_{\rm p}^{\rm rest} \sim 55$~keV, which corresponds to 
an observer-frame $E_{\rm p} \sim 15$ keV. 
This value is consistent with the nondetection of $E_{\rm p}$ in the BAT energy range. 

To date, X-ray afterglows of XRFs have been detected in just a few cases 
(XRF 011030, XRF 020427: \citealt{Bloomea03,Levanea05}; XRF 030723: \citealt{Butlerea04};
XRF 040701: \citealt{Fox04}; XRF 050315: \citealt{Vaughanea06}). 
This is one of the first examples of a well-studied X-ray light curve of an XRF. 
Its main characteristics are not qualitatively different from those of normal GRBs 
(\citealt{Chincaea05,Nousekea05}). 
As observations accumulate, it is becoming clear that these two classes of 
phenomena share many properties, and both have afterglows with similar 
characteristics (\citealt{Sakamotoea05}). 
This is a clue that both types of events may have a common origin and is 
supported by recent evidence that some XRFs are associated with supernovae
(\citealt{Soderbergea04,Bersierea05,Fynboea04}).

\subsection{X-ray flares: evidence for prolonged engine activity\label{grb050406:disc_flares}}

The general behaviour of the afterglow of XRF 050406 is a typical one. 
The observed X-ray photon index ($\Gamma_{\rm X} = 2.1$) 
is common among X-ray afterglows (\citealt{Chincaea05,dePasqualeea05}). 
The light curve shows a break from a relatively steep decay ($\alpha_1=1.58$) 
to a flatter one ($\alpha_2=0.50$). Its overall shape is similar to the one
typically observed by the XRT (\citealt{Chincaea05,Nousekea05}), even though
the initial slope is less steep than average.

The most striking characteristic of this burst is the strong flare in its X-ray 
light curve, a feature which had never been detected by Swift before and had been previously  
observed in very few GRBs  
(GRB~970508, \citealt{piroea99};  GRB~011121 and GRB~011211, \citealt{piroea05}). 
The fluence of the flare is $\sim 1.4 \times 10^{-8}$ erg cm$^{-2}$ in the 
0.2--10 keV band, which amounts to  $\sim 14$\,\% of the observed (15--350 keV band) 
prompt fluence. 
A better estimate of the flare-to-prompt energy ratio would require 
the knowledge of the prompt spectral energy distribution (SED). 
Since the actual peak energy of the prompt SED is unknown ($E_{\rm p} < 15$\,keV), 
the extrapolation of the BAT fluence to the XRT band is highly uncertain. 
For plausible values of $E_{\rm p}$, the flare to prompt fluence ratio
is in the 1--10\,\% range. 
The observed rebrightening is by a factor of 6 in flux, presents a peak at 
$t_{\rm peak}=213\pm7$\,s 
and takes place on a very short timescale, with a  
ratio of the characteristic time-scale and the peak time 
$\delta t / t_{\rm peak} \ll 1$.
Both the rising and the falling parts of the afterglow-subtracted flare had very steep slopes,  
$\alpha_{\rm 1,flare} \approx -7$ and $\alpha_{\rm 2,flare} \approx 7$, 
assuming the burst trigger as the time origin.

According to the standard relativistic fireball model, the prompt emission is 
caused by internal shocks within the expanding fireball, while the afterglow is 
produced by the fireball shocking the external medium (external shocks, \citealt{Piran99,ZM04}). 
Available models to explain flares include refreshed shocks (\citealt{Reesea98}),
external shocks with a clumpy medium (\citealt{Lazzatiea02}) and 
angular inhomogeneities in the outflow (\citealt{Fenimoreea99,Nakarea03}).
However, it can be argued (\citealt{Burrowsea05b}, \citealt{Zhangea05}, 
\citealt{Nousekea05}) that such models cannot produce the observed large 
flux variations $\delta F / F_{\rm peak} \gg 1$ in such short timescales 
$\delta t / t_{\rm peak} \ll 1$ (\citealt{Iokaea05}).
Similarly, none of the above mechanisms would explain the steep slopes observed in the flare.
External reverse shocks, created when the fireball slows down because of the
interaction with the external medium, are expected to emerge at optical and radio
wavelengths, hence synchrotron self-Compton (SSC) must be invoked to produce emission 
in the X-ray band. This would require carefully balanced conditions (\citealt{Kobayashiea05}). 

\citet{piroea05} suggested that the X-ray flares observed in 
GRB 011121 and GRB 011211 were due to the onset of the afterglow. 
The steep slopes and the short timescale variability can only 
be accounted for within the thick shell scenario (\citealt{saripiran99}). 
\citet{GalliPiro05} successfully modeled XRF 011030 using this model. 
In this scenario, the emission before and after the flare is due to different processes
(prompt tail and afterglow, respectively), hence a discontinuity in the light curve 
is generally expected underlying the flare. This is not the case for XRF 050406,  
where the same component describes the X-ray emission both before and after the flare. 
Even if a fine-tuning may explain this particular event, the lack of a light 
curve break is common to a large fraction of the flares observed by Swift 
(\citealt{flaresproc}).
Therefore, while the explanation of flares in terms of the afterglow onset is 
attractive, it is unlikely to be applicable to the vast majority of 
the X-ray flares seen by XRT. 

A promising mechanism to produce the flare is late internal shocks 
(\citealt{Fanea05,Zhangea05,Kingea05,Pernaea06}),
which implies that the central engine is still
active at $t=213$\,s, even though the prompt emission ended after $t \sim 6$\,s.  
The late-time activity in this case must have a reduced power with respect 
to the prompt emission, as the relative fluences indicate. 
Such a mechanism would naturally explain the steep rise and decay slopes. 
Second, the energy required to power the flare would be much lower than in the 
other scenarios (\citealt{Zhangea05}). 
The indications of spectral evolution throughout the flare further support 
this interpretation. The flare appears to be harder than the underlying afterglow, 
which suggests a distinct origin for this emission. 
Furthermore, there are indications of spectral evolution, which shows the typical 
hard-to-soft pattern. Such a behaviour 
is commonly observed in the prompt emission spikes of GRBs (e.g.\ \citealt{Fordea95}), 
which are produced in internal shocks.  
Further evidence of late engine activity comes from both the flat part of the 
light curve ($\alpha_2 \approx 0.5$, see Sect.\ \ref{grb050406:disc_ag}) 
and possibly by the presence of the late-time bump 
observed at $t \sim 1.7 \times 10^{5}$\,s. 

Following the discovery of a flare in the afterglow of XRF 050406, 
initially reported by \citet{Burrowsea05b}, many others were identified: 
GRB 050502B (\citealt{Falconeea05}), GRB 050724 (\citealt{Barthelmyea05b})
and GRB 050904 (\citealt{Cusumanoea05c}), just to mention a few.  
At the time of writing (2005 Oct), $\sim 50$\,\% of the bursts detected by XRT 
which were immediately re-pointed towards showed flares, 
making flaring quite a common behaviour. 
Furthermore, all the characteristics of the XRF 050406 flares 
have now been observed in most flaring GRBs 
(see \citealt{flaresproc} for a recent review). 
For example, highly significant spectral evolution throughout the flare 
has been reported in GRB 050502B (which was the brightest observed so far) 
and GRB 050724. 
In several cases the flares present large amplitudes and occur on short timescales.
Furthermore, several flares are often observed in the same event, 
at times ranging from $\sim 100$\,s to $10^4$--$10^5$\,s after the burst.
Finally, in most cases the afterglow is clearly present {\it before} the onset 
of the flare, and has consistent decay slope and flux levels with after the flare. 
The present case shows that flares are present both in XRFs and in GRBs.
Since flares are likely tied to the central engine activity, this finding 
further supports the idea that a similar mechanism is at work for both kind 
of events (\citealt{Fanea05}).

\subsection{The X-ray afterglow light curve\label{grb050406:disc_ag}}

The prompt reaction of Swift has allowed us to observe the X-ray 
light curves of GRB afterglows starting from a few tens of seconds 
after the burst explosion. 
In most cases the X-ray light curves are characterized by an initial 
steep decay (up to $\sim 500$\,s) followed by a shallow decay, and then by a 
steeper decay with a second break normally occurring at a few thousand 
seconds later (\citealt{Chincaea05,Nousekea05}). 
The early steep decay seen in the X-ray light curve can be explained as 
the tail of the prompt emission  (however, see \citealt{Panaitescuea05}). 
The few cases where the XRT light curve 
lies well above the extrapolation of the prompt emission into the X-ray band 
can be explained either by a strong spectral 
evolution or by an X-ray flare with the maximum 
located before the XRT observation (\citealt{Tagliaferriea05}). 
There are other instances where the first steep decay is not observed at all
(e.g.\ \citealt{Campanaea05}). 

In the case of XRF 050406, however, the initial slope is shallower than the 
steep values $3 \la \alpha \la 5$ observed in other early afterglows 
(\citealt{Tagliaferriea05}). Moreover, the curvature relation 
$\alpha=\beta+2$ (\citealt{Kumarea00,Dermer04}) is  not 
satisfied, even after taking into account the effects pointed out by 
\citet{Zhangea05} that would alter such relation. 
Therefore, we also investigate whether the initial decline seen in XRF 050406 
is consistent with afterglow emission. 
Comparison of spectral indices and temporal decay slopes with 
theoretical relativistic fireball models (e.g.\ Table~2 in \citealt{Zhangea05}) 
indicates that the first decay index $\alpha_1=1.58\pm0.17$ and 
energy index $\beta=1.1\pm0.3$ rule out fast cooling models 
(for which the injection frequency $\nu_{m}$ exceeds the cooling frequency $\nu_{c}$) 
for $\nu < \nu_{\rm m}$. 
For $\nu > \nu_{\rm m}$, the $\alpha(\beta)=(3\beta-1)/2$ closure relation is 
satisfied within the errors and an electron power-law distribution 
index $p \approx 2.5$ is obtained.
The same relation holds for the slow cooling regime (where $\nu_{c} > \nu_{m}$)
for $\nu > \nu_{\rm c}$  (both wind and ISM). In this case a consistent solution is 
also found for $\nu_{\rm m} < \nu < \nu_{\rm c}$, although with a large $p \approx 3$.
The ISM environment is favoured on the basis of a better satisfied closure relation. 
In conclusion, the spectral indices and temporal decay slopes of the first part of the 
X-ray curve can be interpreted in terms of relativistic fireball models,
even though the large uncertainties associated with the slopes do not allow us to choose
among the available models. 

An alternative explanation for the initial XRT emission is the 
presence of an additional flare which started before the beginning of the XRT observation, 
and of which we only see the decaying part. The superposition of two (and possibly more, 
fainter) flares would then mimic the initial steep power law decay. 
However, this interpretation seems less likely since recent Swift observations 
of X-ray flares within the first several hundred seconds of the prompt emission 
all had temporal decay indices much steeper than the observed XRF 050406 
pre-flare index.

At  $t \sim 4400$ s the XRT light curve breaks to $\alpha_2 \approx 0.5$. 
Such a flat decay cannot be explained in terms of the standard afterglow model. 
The only possibility would be to observe, in the fast cooling regime, the segment with 
$\nu_c < \nu < \nu_m$ (where $\alpha = 0.25$ is expected, marginally consistent 
with the observed value). However, the fast cooling regime is expected to end much 
earlier. To maintain the observed decay unbroken up to $\sim 10^6$\,s, 
large values of the equipartition parameters $\varepsilon_e$ and $\varepsilon_B$ 
or of the Compton parameter would be required. 
We consider this possibility quite unlikely. 
Another possibility is that the angular energy profile of the fireball is not trivial 
(a structured jet), so that emission coming from the (brighter) wings of the jet may 
increase the observed flux as the fireball Lorentz factor decreases (\citealt{Panaitescuea05}). 

An interesting explanation for the shallow-decay phase is injection of new 
energy into the fireball through refreshed shocks (\citealt{Sariea00,ZM01}). 
For this to happen, the energy 
release inside refreshed shocks must be sizeable, since the whole fireball dynamics 
has to be modified. Assuming an energy injection rate $\dot{E} \propto t^{-q}$, we find 
$q$ in the range 0 to 0.5 depending on the model details  (\citealt{Zhangea05}). 
In this model, the initial part of the XRT afterglow light curve can be due to standard 
afterglow emission only if the fireball evolution is not influenced at these stages. 
Indeed, the energy supply provided by refreshed shocks is steadily growing, 
and at the beginning it cannot alter the fireball dynamics. 
In this case, the break would identify the time when the new, injected 
energy is comparable to the fireball energy. 
On the contrary, if the first XRT phase were due to late engine activity, 
then the energy injection could have begun much earlier 
and its emission would have been masked. 

Integration of the light curve from the onset of the flat slope phase yields 
$\mathcal{F} \approx 3 \times 10^{-8} (t_{\rm end}/7.6 \times 10^{5} ~ {\rm s})^{0.5}$ 
erg cm$^{-2}$, where $t_{\rm end}$ is the time at which the shallow phase ends, for 
which we can only set a lower limit. 
We note that this depends weakly on the onset time of the shallow phase, therefore 
the calculated fluence is correct in both presented scenarios.
For comparison, the amount of energy released during the steep phase of the light 
curve (excluding the flare) is 
$\mathcal{F} \approx 2 \times 10^{-8} (t_{\rm start}/ 100~{\rm s})^{-0.6}$ 
erg cm$^{-2}$. 
We note that the shallow phase lasts a considerable time. 

\citet{Zhangea05} propose three explanations for the energy injection mechanism.
In the impulsive case (\citealt{Sariea00}), 
the central engine ejects material with a wide distribution of Lorentz factors.
In this case, slower moving shells will catch the fireball at a later time.
We can estimate the minimum Lorentz factor as 
$\Gamma_{\rm min} \la 2 (E_{\rm iso,50} /n_0)^{1/8} (1+z)^{3/8}$, where 
$E_{\rm iso}=E_{\rm iso,50} \times 10^{50}$ erg is the isotropic-equivalent energy, 
and $n_0$ is the external medium particle density in units of cm$^{-3}$. 
This implies that the acceleration process works from ultra- to mildly-relativistic
velocities. 
Within the putative Poyinting flux scenario (\citealt{Zhangea05b}), 
the energy supply is provided by the transfer of 
magnetic energy to the fireball, and the time at which the injection stops is related to
the ratio $\sigma$ of the electromagnetic to baryonic kinetic energy. 
If this scenario is correct, we can infer a lower limit 
of $\sigma = (t_{\rm end}/t_{\rm start})^{1-q} > 10$--$100$,
for $q=0.5$--0, where $t_{\rm start} < t_{\rm b}$ is the start time of injection. 
Therefore, after the end of the energy transfer phase, the energy of the blast-wave would 
be increased by a comparable factor.
In the third scenario (the prolonged energy output by the central engine, \citealt{ZM01}),  
the end of the injection phase is simply the end of the engine activity.
In this case, this activity produces a large amount of energy, particularly so since 
the radiative efficiency may be lower during the late afterglow than during the prompt 
emission, as is generally the case. 
This was previously noticed by \citet{Nousekea05} in a sample of several Swift GRBs.  

The monitoring of XRF 050406 was discontinued 22 days after the trigger. 
By then, the source was no longer detectable and only a 
3-$\sigma$ upper limit could be drawn at  $\approx 3.6 \times 10^{-4}$ counts s$^{-1}$. 
In order for the afterglow energy not to diverge, a further, late break is necessary. 
One interesting possibility is that this may be due to seeing the edge of the jet. 
A steepening in the light curve is expected when the fireball Lorentz factor becomes 
comparable to the inverse of the jet half-opening angle. Such a late break is not 
unexpected for an XRF. The few XRFs with known redshift 
(\citealt{Soderbergea04,Bersierea05,Fynboea04}) 
have a very low isotropic-energy release, and this may be at least in part accommodated 
if they have very wide jets. 
This picture is consistent with the result found by 
Frail et al.\ (2001; see also \citealt{Ghirlandaea04}), 
who found that low-energy GRBs tend to have wider opening angles. 
Using the standard formalism (\citealt{Rhoads99,Sariea99}), the jet half-opening angle is 
$\vartheta_{\rm j} = 16 \, t_{\rm j,6}^{3/8} n_0^{1/8} (\eta/0.2)^{1/8} E_{\rm iso,50}^{-1/8}$ deg,
where $t_{\rm j}=t_{\rm j,6}\times 10^6$\,s is the jet break time and  $\eta$ is the 
burst radiative efficiency. 
Therefore, using our lower limit on the jet break time $t_{\rm j} \ga 10^6$\,s, 
we can infer a lower limit on the jet half-opening angle of 16 deg.
This value is at the high end of the distribution of jet angles (\citealt{Bloomea03b}).

	\section{Summary and conclusions\label{grb050406:conclusions}} 

XRF 050406 is classified as an X-ray flash, with fluence 
$\sim 1 \times 10^{-7}$ erg cm$^{-2}$ (15--350 keV), 
a soft spectrum ($\Gamma_\gamma=2.65$), no significant
flux above $\sim 50$ keV and a peak energy $E_{\rm p} < 15$ keV. 
Its main characteristics are however not qualitatively different from those of 
normal GRBs. As observations accumulate, it becomes clear that these two 
classes of phenomena share many properties, and both have afterglows with similar 
characteristics. This is a clue that both events may have a common origin. 

XRF 050406 is the first Swift-detected burst that showed a flare in its 
X-ray light curve, a feature  now found in $\sim 50$\,\% of the XRT afterglows. 
The flare peaked at $\sim 210$ s after the BAT trigger 
($\sim 61$ s in the rest frame). 
The best fit of the afterglow decay is obtained with a broken power law with  
$\alpha_1=1.58\pm0.17$,  $\alpha_2=0.50^{+0.14}_{-0.13}$, 
and a break at $\sim 4400$ s after the BAT trigger. 
The mean photon index is $\Gamma_{\rm X} = 2.1\pm0.3$.
During the X-ray flare a flux variation of $\delta F / F_{\rm peak} \sim 6$ in 
a timescale $\delta t / t_{\rm peak} \ll 1$ is observed, and its measured fluence 
in the 0.2--10 keV band is $\sim 1.4 \times 10^{-8}$ erg cm$^{-2}$ 
[$(2.0 \pm 1.4) \times 10^{50}$ erg],
which corresponds to 1--15\% of the prompt fluence. 
We argued that the flare-producing mechanism is late internal shocks,
which implies that the central engine is still active at $t \sim 210$\,s,
though with a reduced power with respect to the prompt emission. 
We showed possible indications of spectral variations during the flare, and a 
flattening of the X-ray light curve after  $t \sim 4400$\,s in 
support of continued central engine activity at late times. 

Since XRF 050406 was observed, flares have been detected by XRT in both 
X-ray flashes and normal GRBs, indicating that flares are linked to some 
common properties of both kinds of bursts, and probably tied to their 
central engine.

\begin{acknowledgements}
This work is supported at OAB by ASI grant I/R/039/04, 
at Penn State by NASA contract NAS5-00136 and 
at the University of Leicester by PPARC. 
We gratefully acknowledge the contributions of dozens of members of the XRT and UVOT team at
OAB, PSU, UL, GSFC, ASDC, and MSSL and our subcontractors, who helped make this instrument possible.
\end{acknowledgements}

\begin{table*}[H]
 \begin{center}
 \caption{Observation log of XRF 050406.}
 \label{grb050406:tab_obs}
 \begin{tabular}{cccccc}
 \hline
 \hline
 \noalign{\smallskip}
Sequence         & Obs/Mode     & Start time  (UT)             & End time   (UT)                 & 
Exposure$^{\mathrm{a}}$ & Time since trigger   \\
                 &              & (yyyy-mm-dd hh:mm:ss)           & (yyyy-mm-dd hh:mm:ss)        &(s)  & (s)       \\
 \noalign{\smallskip} 
 \hline 
 \noalign{\smallskip} 
00113872000	&       BAT     &      2005-04-06  15:53:48    &       2005-04-08  06:32:39    &       ...     &      $-$300     \\
00113872000	&	XRT/IM   &      2005-04-06  16:00:12    &       2005-04-06  16:00:14    &       2.5     &       84      \\
00113872000	&	XRT/PuPD &	2005-04-06  16:00:17	&	2005-04-08  00:04:58	&	92	&	90	\\
00113872000	&	XRT/LrPD &	2005-04-06  16:00:18	&	2005-04-08  09:33:32	&	681	&	91	\\
00113872000	&	XRT/WT	&	2005-04-06  16:00:20	&	2005-04-08  09:38:58	&	3728	&	92	\\
00113872000	&	XRT/PC	&	2005-04-06  16:00:26	&	2005-04-08  09:49:03	&	49939	&	99	\\
00113872001	&	XRT/PC	&	2005-04-08  09:49:39	&	2005-04-08  22:51:57	&	7431	&	150652	\\
00113872002	&	XRT/PC	&	2005-04-09  00:14:14	&	2005-04-11  23:09:57	&	12558	&	202527	\\
00113872005	&	XRT/PC	&	2005-04-12  00:36:42	&	2005-04-12  23:20:12	&	3912	&	463074	\\
00113872006	&	XRT/PC	&	2005-04-13  00:29:18	&	2005-04-13  23:26:58	&	11291	&	549031	\\
00113872008	&	XRT/PC	&	2005-04-16  00:48:33	&	2005-04-18  23:43:57	&	35937	&	809386	\\
00113872009	&	XRT/PC	&	2005-04-20  01:13:27	&	2005-04-20  23:59:57	&	17937	&	1156479	\\
00113872010	&	XRT/PC	&	2005-04-21  01:19:44	&	2005-04-21  23:59:58	&	12546	&	1243256	\\
00113872011	&	XRT/PC	&	2005-04-22  00:05:06	&	2005-04-22  22:48:57	&	11526	&	1325178	\\
  \noalign{\smallskip}
  \hline
  \end{tabular}
  \end{center}
  \begin{list}{}{}
  \item[$^{\mathrm{a}}$] The exposure time is spread over several snapshots (single continuous pointing at the target) 
	during each observation (with the exclusion of BAT and XRT/IM data).
  \end{list}
 \end{table*}

 \begin{table*} 	
 \begin{center} 	
 \caption{Spectral fit results.} 	
 \label{grb050406:tab_specfits} 	
 \begin{tabular}{lllllll} 
 \hline 
 \hline 
 \noalign{\smallskip} 
  Spectrum	& Photon index	 		& 	$N_{\rm H}$	 & $\chi^{2}_{\rm red}$ (d.o.f.) & C-stat (\%)$^{\mathrm{a}}$  
		& Start time                    & End time  \\
 		& 	 	 		& ($10^{20}$ cm$^{-2}$)  & 	 & & (s since $T_0$) 	&   (s since $T_0$) 	   \\
 \noalign{\smallskip} 
 \hline 
 \noalign{\smallskip} 
BAT total		&$2.63_{-0.36}^{+0.42}$	&	...	& 1.4 (56) &...	&  -2.560	& 4.160   \\
BAT $T_{90}$		&$2.64_{-0.38}^{+0.46}$	&	...	& 1.3 (56) &...	&  -2.432	& 3.648   \\
BAT $T_{50}$		&$2.65_{-0.48}^{+0.61}$	&	...	& 1.0 (56) &...	&  -0.064	& 2.048   \\
BAT peak		&$2.65_{-0.60}^{+0.82}$	&	...	& 0.9 (56) &...	&   0.064	& 1.024   \\
 \noalign{\smallskip} 
 \hline 
 \noalign{\smallskip} 
XRT    WT 		&$2.11_{-0.28}^{+0.31}$ & 2.8$^{\mathrm{b}}$  	& 1.0 (6) & ... &  92	& 596   \\
XRT WT+PC 		&$2.12_{-0.23}^{+0.25}$ & 2.8$^{\mathrm{b}}$ 	& 1.0 (8) & ... &  92	& 599   \\
XRT    PC 		&$2.13_{-0.19}^{+0.44}$ & 2.8$^{\mathrm{b}}$  	& ... 	  & 220.7 (58.0) & 99 	& 599   \\
XRT    PC$^{\mathrm{c}}$ &$2.06_{-0.24}^{+0.24}$   & 2.8$^{\mathrm{b}}$	& ...     & 270.7 (67.1) & 599 	& 18308   \\
  \noalign{\smallskip}
  \hline
  \end{tabular}
  \end{center}
  \begin{list}{}{}
  \item[$^{\mathrm{a}}$] Cash statistic (C-stat) and percentage of Monte Carlo realizations 
			 that had statistic $<$ C-stat. We performed $10^4$ simulations. 
  \item[$^{\mathrm{b}}$] Fixed to the Galactic value.
  \item[$^{\mathrm{c}}$] Snapshots 1 through 4, with the exclusion of the first $\sim 600$\,s 
			(non piled-up data). 
   \end{list}
  \end{table*} 

 \begin{table*}
 \begin{center}
 \caption{Light curve fit results$^{\mathrm{a}}$.}
 \label{grb050406:tab_lcvfits}
 \begin{tabular}{llllll}
 \hline
 \hline
 \noalign{\smallskip}
 Model parameters &  Simple power law     		&  Broken power law     		& Smoothly-joined power laws 		& Broken power law+Gaussian\\
 		  &  excluding flare      		&  excluding flare 			& excluding flare 		& full data set  \\
 \noalign{\smallskip}
 \hline
 \noalign{\smallskip}
$\alpha_1$	& $1.41^{+0.22}_{-0.24}$		& $1.58^{+0.18}_{-0.16}$		& $1.73^{+0.40}_{-0.24}$ 	& $1.58\pm0.17$	\\
$t_{\rm b}$ (s)	& -					& $(4.19^{+6.17}_{-0.36})\times 10^{3}$	& $(3.61^{+1.36}_{-1.03})\times 10^{3}$	& $(4.36^{+6.23}_{-0.53})\times 10^{3}$\\
$\alpha_2$	& -					& $0.50\pm0.14$			        & $0.42_{-0.12}^{+0.11}$	& 
$0.50^{+0.13}_{-0.14}$\\
Gaussian centre (s)	& -					& -					& - & $211.1^{+5.4}_{-4.4}$\\
Gaussian width	(s)	& -					& -					& - & $17.9^{+12.3}_{-4.6}$\\
$\chi^2_{\rm red}$ & 4.32					& 1.20					& 1.29 & 1.58\\
d.o.f.		& 12					& 10					& 10 & 17\\
  \noalign{\smallskip}
  \hline
  \end{tabular}
  \end{center}
  \begin{list}{}{}
  \item[$^{\mathrm{a}}$   We follow the notation $F(t) \propto t^{-\alpha}$.]
  \end{list}
  \end{table*}

\end{document}